\documentclass[10pt,conference]{IEEEtran}
\IEEEoverridecommandlockouts
\usepackage{amsmath, amsfonts, amssymb, amsthm}
\usepackage{algpseudocode}
\usepackage{algorithm}
\usepackage{algorithmicx}
\usepackage{array}
\usepackage{textcomp}
\usepackage{stfloats}
\usepackage{url}
\usepackage{verbatim}
\usepackage{graphicx}
\usepackage{cite}
\usepackage{enumitem}
\usepackage[caption=false,font=footnotesize]{subfig}
\usepackage{xcolor}
\begin{document}

\title{Optimization for Massive 3D-RIS Deployment: A Generative Diffusion Model-Based Approach }
\author{
\IEEEauthorblockN{Kaining Wang\IEEEauthorrefmark{1}, Bo Yang\IEEEauthorrefmark{1}, Zhiwen Yu\IEEEauthorrefmark{1}\IEEEauthorrefmark{2},  Xuelin Cao\IEEEauthorrefmark{3}, 
M\'erouane Debbah\IEEEauthorrefmark{4},  Chau Yuen\IEEEauthorrefmark{5}} 
\IEEEauthorblockA{\IEEEauthorrefmark{1}School of Computer Science, Northwestern Polytechnical University, Xi'an, Shaanxi, 710129, China} 
\IEEEauthorblockA{\IEEEauthorrefmark{2}Harbin Engineering University, Harbin, Heilongjiang, 150001, China} 
\IEEEauthorblockA{\IEEEauthorrefmark{3}School of Cyber Engineering, Xidian University, Xi'an, Shaanxi, 710071, China}
\IEEEauthorblockA{\IEEEauthorrefmark{4}Center for 6G Technology, Khalifa University of Science and Technology, P O Box 127788, Abu Dhabi, UAE}
\IEEEauthorblockA{\IEEEauthorrefmark{5}School of Electrical and Electronics Engineering, Nanyang Technological University, Singapore}
}

\maketitle

\begin{abstract}
Reconfigurable Intelligent Surfaces (RISs) transform the wireless environment by modifying the amplitude, phase, and polarization of incoming waves, significantly improving coverage performance. Notably, optimizing the deployment of RISs becomes vital, but existing optimization methods face challenges such as high computational complexity, limited adaptability to changing environments, and a tendency to converge on local optima.
In this paper, we propose to optimize the deployment of large-scale 3D RISs using a diffusion model based on probabilistic generative learning. We begin by dividing the target area into fixed grids, with each grid corresponding to a potential deployment location. Then, a multi-RIS deployment optimization problem is formulated, which is difficult to solve directly.
By treating RIS deployment as a conditional generation task, the well-trained diffusion model can generate the distribution of deployment strategies, and thus, the optimal deployment strategy can be obtained by sampling from this distribution. Simulation results demonstrate that the proposed diffusion-based method outperforms traditional benchmark approaches in terms of exceed ratio and generalization.


\end{abstract}


\section{Introduction}

The rapid growth of the Internet of Things (IoT) and machine-type communications requires wireless networks that can provide reliable, large-scale connectivity in dynamic and obstacle-rich environments. Recently, reconfigurable intelligent surfaces (RISs) have emerged as a promising solution to tackle these challenges. 
By adjusting the electromagnetic properties of nearly passive reflective elements, RIS can manipulate incoming signals to boost desired transmissions, suppress interference, and expand coverage, all without increasing power consumption at the transmitter or receiver \cite{RISapp1,RISapp2,RISapp3}.

Despite the rapid advancements in RIS-enabled communications, the development of practical deployment strategies remains a critical issue that has not been thoroughly explored. The placement of RISs significantly affects the reflected signals, which, in turn, influences the received signal-to-noise ratio (SNR) and overall coverage performance. Recently, there has been an increasing amount of research focusing on how different RIS placements impact communication performance.  For example, Ren \textit{et al.} \cite{deploymentDistance1}  experimentally analyzed the optimal distances between RISs, base stations (BS), and users in general wireless systems, establishing empirical relationships between geometric configurations and signal enhancement. The analysis was further expanded to include relay systems by developing energy-efficient deployment strategies that aim to maximize coverage in relay-aided architectures  \cite{deployment2}.


In challenging propagation environments, Deb \textit{\textit{et al.}} \cite{deployment4} proposed a strategy for deploying RIS to enhance millimeter-wave device-to-device (D2D) communications. This strategy is designed to circumvent static obstacles and restore blocked communication links. To improve coverage in large-scale networks, Ma \textit{\textit{et al.}} \cite{deployment6} introduced a graph-based approach for the joint deployment of BS and RIS. Their method partitions the coverage area into cells that contain potential locations for placement. Additionally, \cite{deployment7} proposed both graph-based and multi-RIS frameworks to jointly optimize the placement of BS and RIS, aiming to enhance coverage while balancing cost and performance. While these studies encompass a range of scenarios from single-RIS to multi-RIS systems, they are limited by their reliance on predefined candidate locations. Furthermore, they predominantly focus on two-dimensional deployments. This focus restricts their applicability to real-world situations, where environments are inherently three-dimensional, and optimal RIS placement may not align with predefined locations.


Recent research has gradually advanced the optimization of RIS deployment in three-dimensional (3D) environments. For instance, Fu \textit{et al.} \cite{deployment8} formulated a multi-RIS deployment problem by discretizing the service area into grids. They utilized a branch-and-bound (BB) algorithm to choose from a predefined set of candidate sites while simultaneously optimizing the height, orientation, and number of tiles for each RIS. This method showed significant performance improvements and reduced deployment costs compared to traditional strategies. However, it is based on the assumption that the target area contains a fixed set of candidate deployment sites. Furthermore, the grid-based discretization introduces approximation errors in channel power evaluation, resulting in prohibitively complex computational requirements for large-scale deployments.


The process of deploying RISs can be viewed as finding an optimal solution to an optimization problem. Many studies have explored various learning-based approaches to address complex optimization challenges in wireless networks. Traditional supervised methods, such as multi-task feedforward neural networks (MTFNN) \cite{MTFNN}, provide effective mappings from environmental features to deployment strategies. However, these methods generally require large amounts of labeled data. On the other hand, deep reinforcement learning (DRL) \cite{DRL}, particularly through techniques like proximal policy optimization (PPO), has been widely studied for adaptive deployment. Nonetheless, its performance is often sensitive to the design of reward systems and training stability. Notably, diffusion models have emerged as powerful generative optimizers, and several studies have investigated the potential of diffusion for generating optimal solutions \cite{survey, diffusion3, GDSG, jsac}. 

Driven by these advancements, we approach the RIS deployment problem as a conditional generation task. In this framework, the model is trained to generate the coordinates and heights of the RIS based on the positions of the BS, user locations, and information about obstacles. Extensive experiments are conducted across various network scenarios to validate the effectiveness of the proposed framework.

\section{System Model and Problem Formulation}

\begin{figure*}[htbp]
    \centering
    \captionsetup{font={small}}
    \subfloat[]{\includegraphics[width=0.5\textwidth]{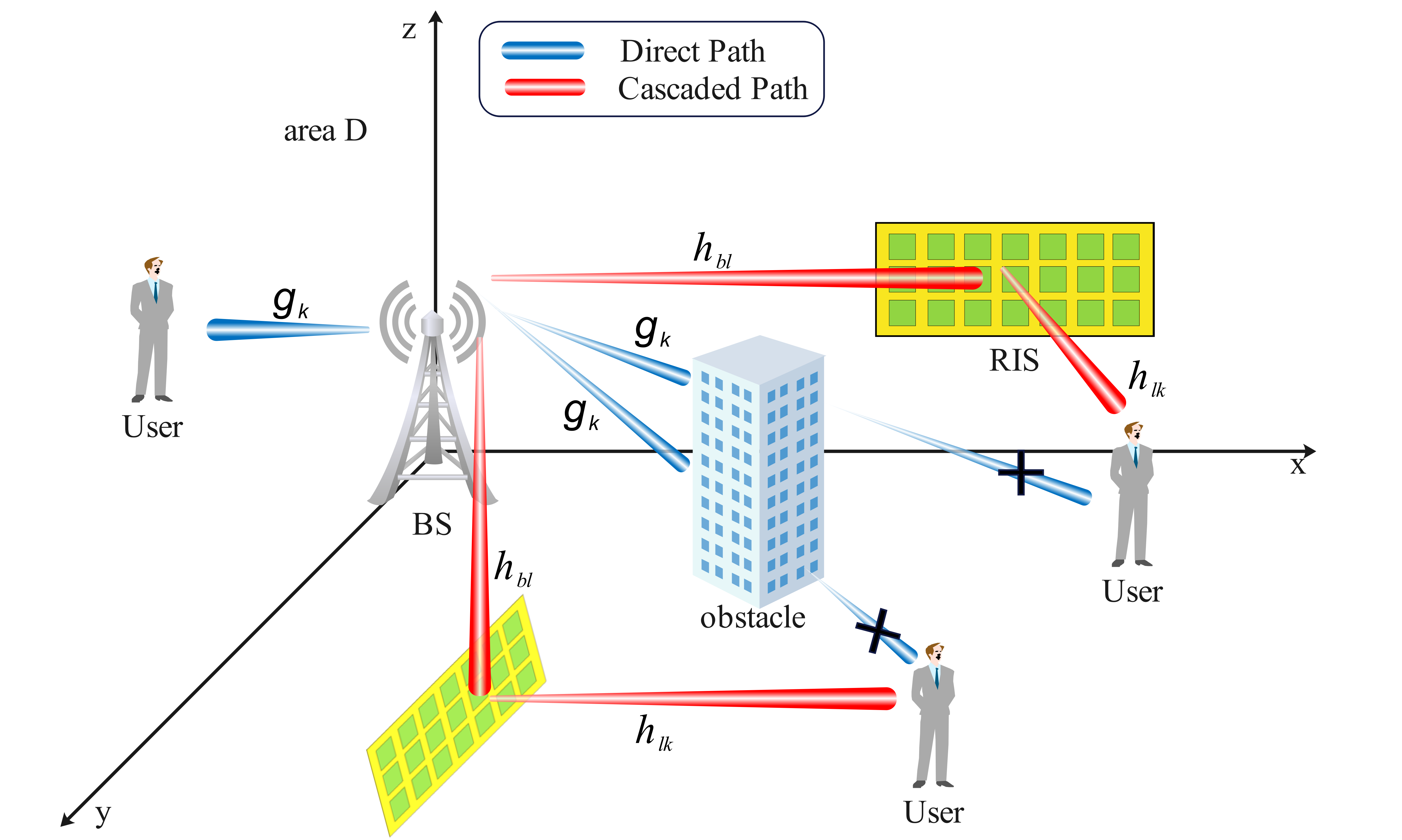}}
    \hfill 
    \subfloat[]{\includegraphics[width=0.5\textwidth]{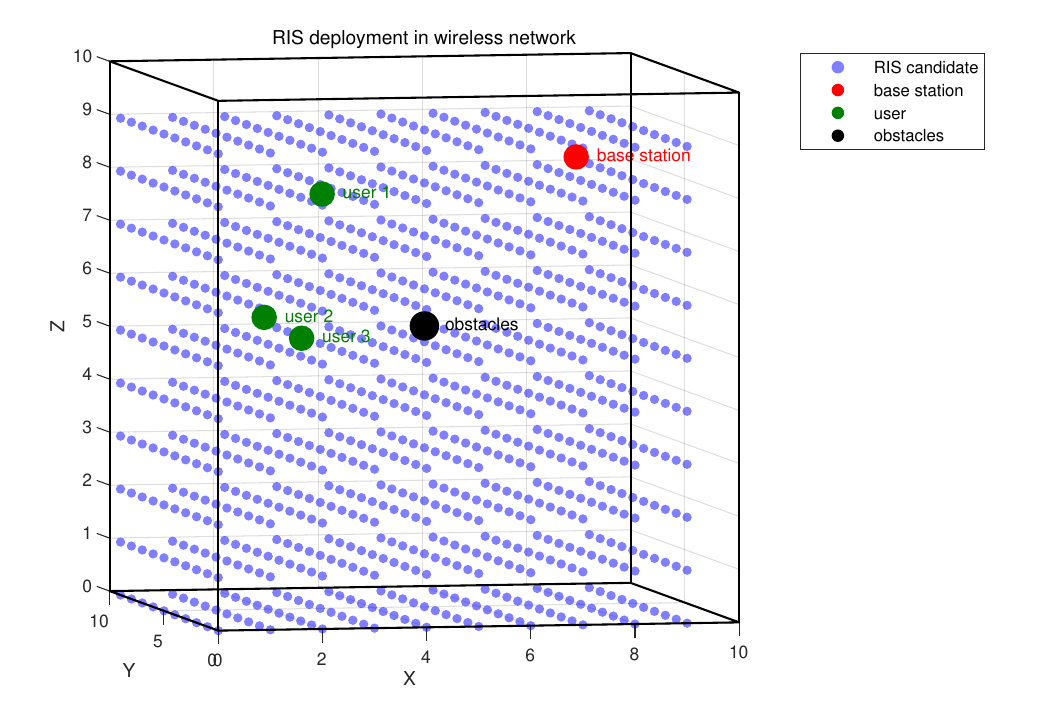}}
    \caption{(a) System model of a RIS-assisted wireless network in which the BS communicates with multiple users despite the presence of obstacles. RISs are deployed within region D to create cascaded reflection links (shown in red) in addition to direct paths (shown in blue). (b) A three-dimensional RIS deployment scenario. The BS, multiple users, and obstacles are located within region D, which is divided into candidate grid points (shown in blue) for the deployment of RISs. }
    \label{fig:experiment}
\end{figure*}

\subsection{System Model}
As shown in Fig.~\ref{fig:experiment}(a), we are examining a wireless communication system that consists of one BS, $K$ users, and $L$ RISs. These RISs are deployed within the area \( D \) to create virtual line-of-sight (LoS) paths. Since any location in three-dimensional space could potentially be an optimal site for deploying RIS, we discretize the area into non-overlapping grid cells to make the problem more tractable. Moreover, each RIS is made up of \(N\) passive reflecting elements that can adjust the phase shifts of incoming signals to either bypass obstacles or coherently combine signals at the receiver.


As illustrated in Fig.~\ref{fig:experiment}(b), the 3D service region $D$ is discretized into $M$ potential grid points, each grid point is denoted as $\xi_i=(x_i,y_i,z_i),i=1,2,\ldots,M$, where $x_i,y_i,z_i$ indicates the three-dimensional coordinates of $\xi_i$, respectively. RIS deployment is represented by a binary selection vector, $\mathbf{s}=[s_1,\ldots,s_M]\in\{0,1\}^M$, where $s_i=1$ indicates that an RIS is placed at grid point $\xi_i$. Otherwise, $s_i=0$ holds. The total number of deployed RISs meets the constraint $\sum_{i=1}^{M} s_i=L$, and $L \leq M$ usually holds. The set of active deployment indices is denoted as ${\mathcal{S}}=\{i \mid s_i=1\}$. Therefore, the position of the $l$th deployed RIS is given by $\chi_l = \xi_{i} s_i, \forall i \in [1,M]$.



\subsection{Signal Model}
The considered propagation model encompasses both the direct BS–user path and the cascaded BS–RIS–user path, accounting for large-scale path loss and small-scale fading effects. The signal received by the $k$th user depends on the existence of a LoS path to the BS.

Let $h_{k}$ denote the small-scale fading of the BS–user-$k$ link, the path loss is expressed as
\begin{equation}
\eta_{d} = Cd_k^{-\alpha},
\end{equation} 
where $d_k$ is the distance between BS and the $k$th user, $\alpha$ is the path loss exponent. $C$ denotes the free-space path loss per unit distance, incorporating antenna gains ($G_t,G_r$) and wavelength ($\lambda$) dependence, i.e.,
$
C = \left(\frac{\lambda \sqrt{G_t G_r}}{4\pi}\right)^2.
$

For the $l$th RIS, $l \in \{1,\ldots,L\}$, let $\mathbf{h}_{bl}\in\mathbb{C}^{N}$ and $\mathbf{h}_{l,k}\in\mathbb{C}^{N}$ denote the small-scale channels for the BS–RIS$_l$ and RIS$_l$–user$_k$ links, respectively. The cascaded BS–RIS$_l$–user$_k$ path loss is given by
\begin{equation}
\eta_{r} = C (d_{bl} d_{lk})^{-\alpha},
\end{equation}
where $d_{bl}$ and $d_{lk}$ denote the distances of BS–RIS$_l$ link and RIS$_l$–user$_k$ link, respectively. 

Each RIS$_l$ is composed of $N$ passive reflecting elements, where each element introduces an adjustable phase shift to the incident signal. The overall phase shifts of RIS$_l$ can be represented by a diagonal matrix:
\begin{equation}
\boldsymbol{\Phi}_l = \mathrm{diag}(e^{j\theta_{l,1}},\ldots,e^{j\theta_{l,N}}), \ \forall l \in [1, L],
\end{equation}
where $\theta_{l,N}$ denotes the phase shift applied by the $n$-th reflecting element of RIS$_l$.

We define a binary indicator, $\delta_k$, to capture the existence of a LoS path for the $k$th user:
\begin{equation}
\delta_k =
\begin{cases}
1, & \text{if a LoS path exists (no obstacles)}, \\
0, & \text{if the LoS path is blocked}.
\end{cases}
\end{equation}

Therefore, the received signal at the $k$th user can be calculated as
\begin{equation}
\zeta_k = \delta_k \cdot \sqrt{\eta_d} \cdot g_k + \sum_{l=1}^{L} \sqrt{\eta_{r}} \cdot \mathbf{h}_{lk}^H \boldsymbol{\Phi}_l \mathbf{h}_{bl},
\end{equation}
where $g_k \sim \mathcal{CN}(0,1)$ is the Rayleigh fading component of the direct link. $\mathbf{h}_{bl}$ and $\mathbf{h}_{lk}$ are the small-scale fading vectors from BS to RIS$_l$ and RIS$_l$ to the $k$th user. $\boldsymbol{\Phi}_l = \mathrm{diag}(e^{j\theta_{l,1}}, \ldots, e^{j\theta_{l,N}})$ indicates the phase shift matrix of RIS$_l$.

The corresponding signal-to-noise ratio (SNR) at the $k$th user is given by
\begin{equation}
\gamma_k = \frac{P |\zeta_k|^2}{\sigma^2  },
\end{equation}
where $P$ is the transmit power, $\sigma^2$ is the noise power.

To evaluate the overall performance of the RIS-assisted multi-user system, we use coverage probability as the key metric. For a specific user \( u \), we determine coverage success by checking if the received SNR exceeds a predefined threshold, denoted as \( \gamma_{\mathrm{th}} \). The overall coverage probability across the domain \( D \) is defined as the average number of users that achieve coverage, i.e.,
\begin{equation}
P_c = \frac{1}{K}\sum_{k-1}^{K} {\rm Pr}[\gamma_k \geq \gamma_{\mathrm{th}}].
\end{equation}

It is evident that $P_c$ signifies the effectiveness of a specific RIS deployment in providing reliable service amidst fading and blockage.

\subsection{Problem Formulation}
We formulate the task of RIS deployment as a 3D optimization problem over a discretized candidate grid, denoted as $\mathbf{\mathcal{P}}$. The objective is to identify the optimal RIS deployment locations that maximize coverage probability, i.e.,
\begin{align}
\mathbf{\mathcal{P}}: & \ \ \max_{\mathbf{s},{\boldsymbol{\Phi}_l}} \quad P_c  \label{Problem}\\
\text{s.t. } 
& s_i \in \{0,1\}, \ \forall i \in [1, M], \tag{\ref{Problem}{a}}\label{Problemb}\\
 &\sum_i^M s_i = L,\tag{\ref{Problem}{b}} \label{Problemc}\\
& \theta \in [0,2\pi), \tag{\ref{Problem}{c}}\label{Problemd}\\
&\|\chi_l - \chi_{l'}\|_2 \ge d_{\min}, \ \forall l \neq l', \tag{\ref{Problem}{d}}\label{Problemf}\\
&x_i \!\in\! [x_{\min}, x_{\max}],y_i \!\in\! [y_{\min}, y_{\max}],z_i \!\in\! [z_{\min}, z_{\max}].\tag{\ref{Problem}{e}}\label{Probleme}
\end{align}


In the optimization problem \(\mathbf{\mathcal{P}}\), the constraints ($\ref{Problem}${a}) and ($\ref{Problem}${b}) restrict the range of variable $s_i$. The range of phase shift is given by ($\ref{Problem}${c}). To prevent colocated deployments, we enforce a minimum spacing requirement between RISs, defined as \eqref{Problemd}. The variable \(x_i\), \(y_i\) and \(z_i\) are constrained to the range \([x_{\min}, x_{\max}]\), \([y_{\min}, y_{\max}]\) and \([z_{\min}, z_{\max}]\), respectively, as explicated in ($\ref{Problem}${e}). 

The binary selection vector \(\mathbf{s}\) creates a discrete and high-dimensional search space that increases combinatorially with the number of potential deployment sites. Additionally, the received SNR is affected nonlinearly by the placement of RISs, due to the interconnected effects of channel propagation, phase shift design, and path loss. Therefore, the problem \(\mathbf{\mathcal{P}}\) represents a nonlinear combinatorial optimization problem that is NP-hard, making it extremely difficult for traditional solvers to handle, particularly in large-scale or dynamic environments. 

\section{Diffusion Model-based Optimization for RIS Deployment}

\begin{figure}
    \centering
    \captionsetup{font={small}}
    \includegraphics[width=1.02\linewidth]{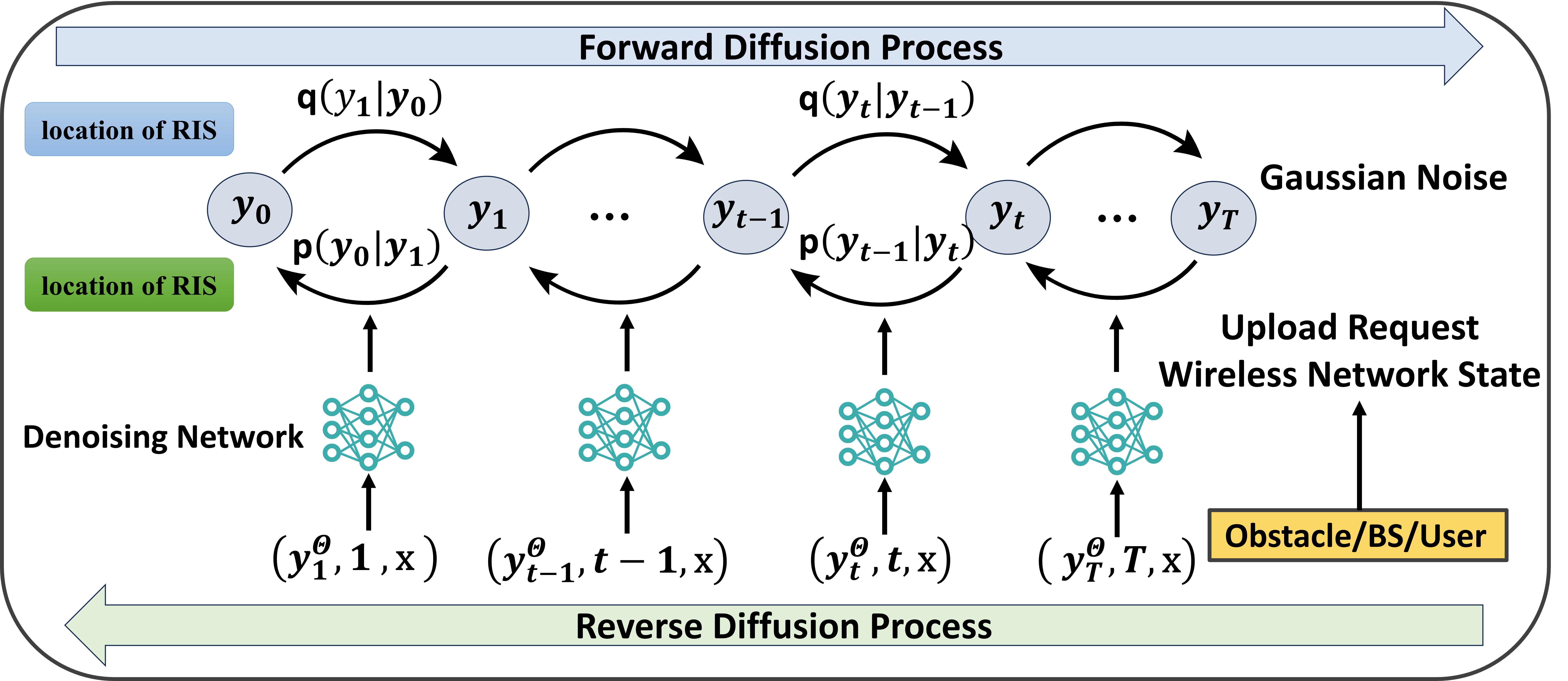}
    \caption{The structure of the diffusion model for RIS deployment optimization.}
    \label{fig:diffusion}
\end{figure}

In this paper, we formulate RIS deployment optimization as a conditional generation task, which is solved using the denoising diffusion probabilistic model (DDPM) \cite{diffusion1}, as illustrated in Fig.~\ref{fig:diffusion}. The proposed approach aims to learn a high-quality distribution of RIS deployment locations conditioned on the positions and spatial layout of BSs, users, and obstacles. By repeatedly sampling from the learned distribution, the model generates deployment strategies that are optimal or nearly optimal.

\subsection{Problem Analysis via Conditional Generation}
The RIS deployment problem involves a high-dimensional, non-convex search space where each deployment decision nonlinearly influences the channels of multiple users. To tackle this challenge, we reformulate the task as a conditional optimization problem. In this problem, the conditional inputs include user locations, the BS position, and obstacle geometry.
The goal is to learn a conditional solution distribution $p_\theta(\mathbf{y}|\mathbf{x})$, where $\mathbf{x}$ encodes the environment and $\mathbf{y}$ denotes the RIS deployment positions.

To capture the environment-conditioned distribution of RIS deployments, we employ a generative diffusion model (GDM), which progressively corrupts data with noise in a forward process and learns to invert this corruption in a reverse process to recover the original samples \cite{diffusion2}. In this context, each training instance is a pair $(\mathbf{x}, \mathbf{y})$, where $\mathbf{x}$ encodes the environment and $\mathbf{y}$ corresponds to the optimal RIS deployment. The target $\mathbf{y}$ is represented either as a multi-hot vector $\mathbf{s}\!\in\!\{0,1\}^M$ with fixed cardinality $L$, or as a set of continuous coordinates $\mathbf{Y}\!=\![\chi_1,\ldots,\chi_L]\!\in\!\mathbb{R}^{L\times 3}$. The conditioning variable $\mathbf{x}$ encodes BS and user locations, obstacle geometry, and visibility maps.  Once the GDM is well-trained, it can infer the distribution of RIS deployment, $p_\theta(\mathbf{y}|\mathbf{x})$, obtaining the optimal coordinates of RISs via sampling \cite{diffusion1}.

\subsection{Diffusion Process Design}
\subsubsection{Forward Process}
The forward (noising) process gradually corrupts the original data over $T$ time steps by adding Gaussian noise, with the noise level at each step controlled by a predefined schedule $\alpha_t$. This process is modeled as a Markov chain:
\begin{equation}
    q(\mathbf{y}_t | \mathbf{y}_{t-1}) \!=\! \mathcal{N}(\mathbf{y}_t; \sqrt{\alpha_t} \mathbf{y}_{t-1}, (1 - \alpha_t)\mathbf{I}), \ t \in [1, T],
\end{equation}
where $\mathbf{y}_{t}$ denotes the noisy deployment representation at step $t$, $\mathbf{y}_{t-1}$ is the previous state, $\mathbf{I}$ is the identity matrix, and the noise is assumed to be independent across dimensions.

By iteratively applying this process, the original deployment sample $\mathbf{y}_0$ is gradually transformed into pure noise $\mathbf{y}_T$. The distribution of $\mathbf{y}_t$ conditioned on $\mathbf{y}_0$ can be expressed in closed form as
\begin{equation}
    q(\mathbf{y}_t | \mathbf{y}_0) = \mathcal{N}(\mathbf{y}_t; \sqrt{\bar{\alpha}_t} \mathbf{y}_0, (1 - \bar{\alpha}_t)\mathbf{I}),
\end{equation}
where $\bar{\alpha}_t = \prod_{i=1}^t \alpha_i$ is the cumulative product of the noise schedule up to step $t$.

\subsubsection{Reverse Process}
The reverse (denoising) process seeks to reconstruct $\mathbf{y}_0$ from $\mathbf{y}_t$ by employing a neural network parameterized by $\theta$, which predicts the noise at each step conditioned on the environment $\mathbf{x}$:
\begin{equation}
    p_\theta(\mathbf{y}_{t-1} | \mathbf{y}_t, \mathbf{x}) = \mathcal{N}(\mathbf{y}_{t-1}; \mu_\theta(\mathbf{y}_t, \mathbf{x}, t), \Sigma_\theta(\mathbf{y}_t, \mathbf{x}, t)),
\end{equation}
where $\boldsymbol{\mu}_\theta$ denotes the model-predicted mean and $\boldsymbol{\Sigma}_\theta$ the predicted variance. The model is trained by minimizing the prediction loss between the estimated and true noise components.

For discrete RIS deployment representations, we employ a multinomial diffusion strategy that progressively perturbs the multi-hot indicator vector \cite{discreteDIFF}. The reverse denoising process is implemented by a neural network and trained using a cross-entropy noise-prediction objective under classifier-free guidance. To ensure the feasibility of the generated deployments, a cardinality constraint is imposed through a penalty term during training, while a Top-$L$ projection is applied at inference to guarantee that exactly $L$ RISs are selected.

\subsubsection{Classifier-Free Guidance}
To ensure that the denoising process is conditioned on the environment $\mathbf{x}$, we adopt classifier-free guidance \cite{ho2022classifier}. This technique combines conditional and unconditional predictions without requiring an auxiliary classifier. The guided noise prediction is defined as
\begin{equation}\label{noiseP}
    \epsilon_{\text{guided}} = (1 + \omega)\epsilon_\theta(\mathbf{y}_t|\mathbf{x}) - \omega\epsilon_\theta(\mathbf{y}_t),
\end{equation}
where $\omega$ is a scalar weight that controls the strength of guidance by the environmental context, and $\epsilon_\theta(\cdot)$ denotes the noise-prediction network used in the reverse denoising process.

\section{Experiments Results and Discussions}
To assess the effectiveness of the proposed diffusion-based framework, we conduct thorough experiments across various deployment scenarios. Initially, we analyze how key parameters of the diffusion model affect deployment performance and its ability to generalize across different environments. Then, we compare our proposed method with several baseline approaches under diverse conditions to highlight its advantages.

\subsection{Experiment Configuration}
\begin{enumerate}
    \item \textbf{Datasets:}
    Training datasets are created through an exhaustive search process that evaluates the optimal placements of RISs. Although this approach is computationally intensive, it ensures the reliability of expert labels. Each sample includes the coordinates of the BS, users, and obstacles, along with the optimal RIS configuration and the corresponding performance metrics. To test generalization, we design nine datasets with varying densities of obstacles and users, ranging from 1 obstacle and 1 user (denoted as 1obs$\_$1user) to 10 obstacles and 10 users (denoted as 10obs$\_$10users). The models are trained on a dataset with 3 obstacles and 3 users (3obs$\_$3users) and are evaluated on unseen scenarios (1obs$\_$3users and 3obs$\_$1user) to assess cross-domain adaptability.

    \item \textbf{Evaluation metrics:}
    Wireless channels are subject to randomness due to fading and variability, which can result in significant fluctuations in absolute SNR values. To address this, we define the exceed ratio as \( E = \frac{\gamma_p}{\gamma_t} \), where \( \gamma_p \) represents the SNR achieved from model-predicted deployments, and \( \gamma_t \) indicates the optimal SNR obtained through exhaustive search. This ratio serves as a normalized measure of performance across different environments, with values close to 1 signifying near-optimal performance.
    
    \item \textbf{Model settings:} 
    The proposed framework utilizes a U-Net network \cite{unet} with classifier-free guidance that is conditioned on the features of the BS, the user, and obstacles. Unless specified otherwise, the number of diffusion steps is fixed at \(T = 20\), the guidance weight is set to \(\omega = 180\), and the unconditional dropout rate is \(p_{\text{uncond}} = 0.1\). For training, the Adam optimizer is used with a learning rate of \(10^{-4}\) along with an exponential moving average to enhance stability.
    
\end{enumerate}

\subsection{Impacts of Model Parameters}
\begin{figure*}
    \centering
    \subfloat[]{\includegraphics[width=0.32\textwidth]{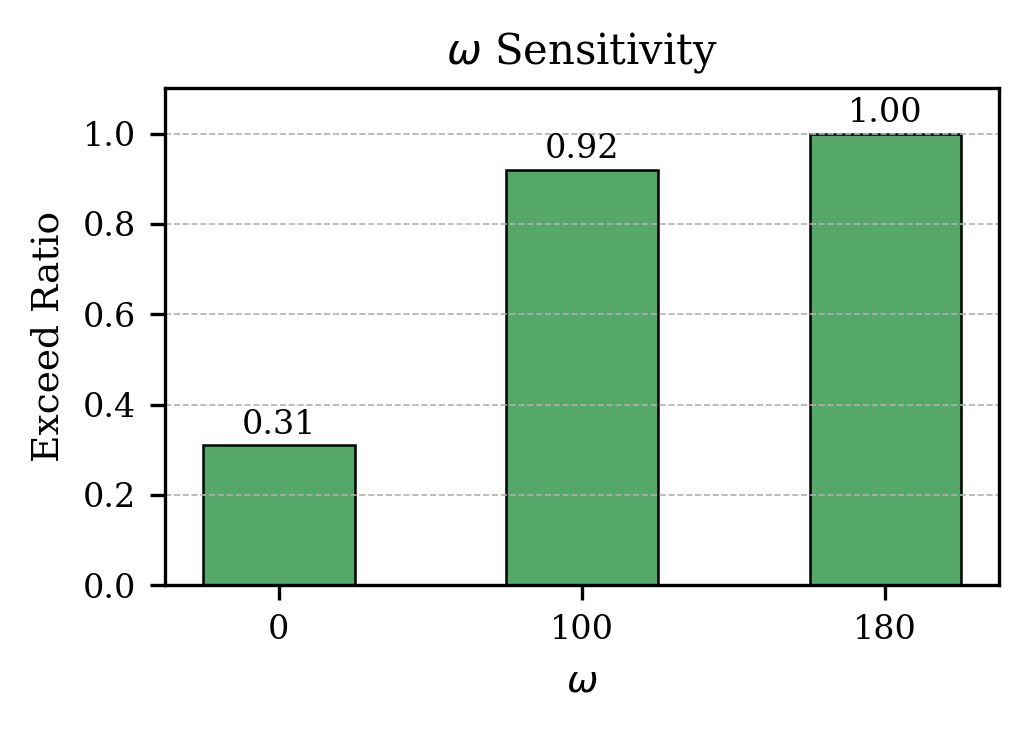}}
    \hfill
    \subfloat[]{\includegraphics[width=0.32\textwidth]{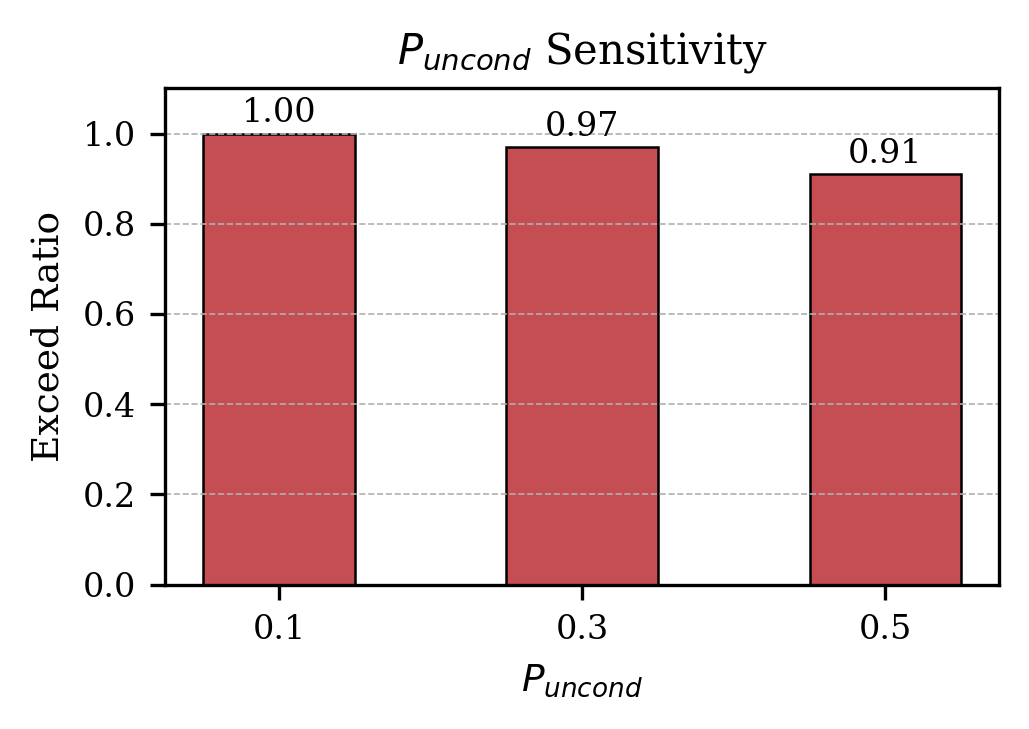}}
    \hfill
    \subfloat[]{\includegraphics[width=0.32\textwidth]{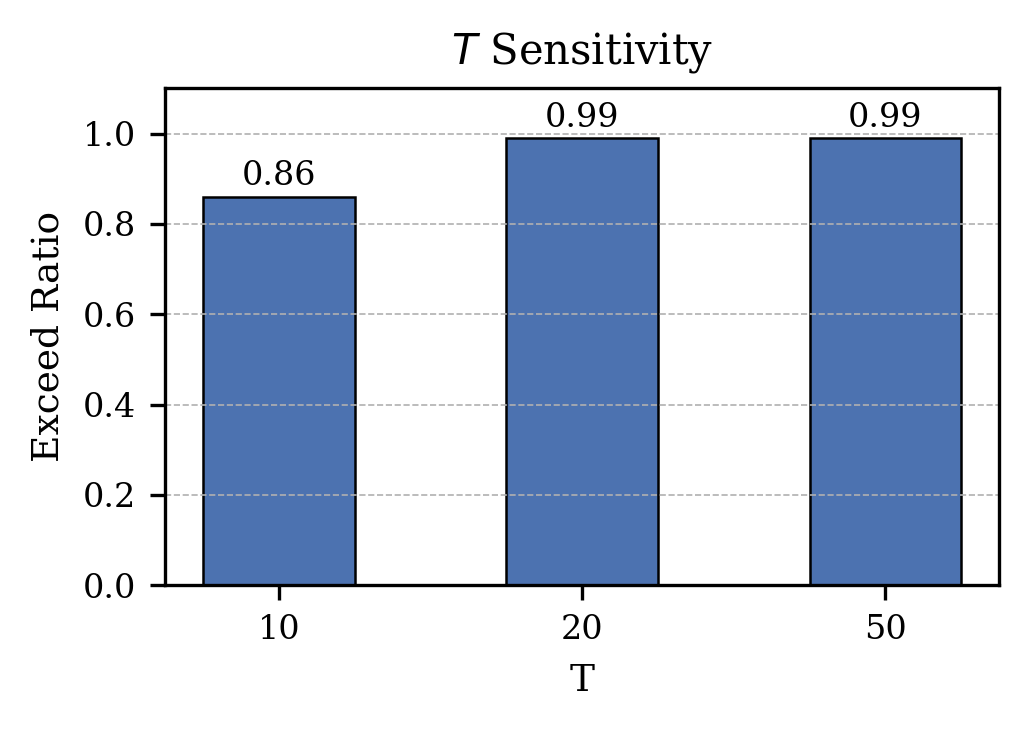}}
    \caption{The sensitivity of the diffusion-based framework. (a)Impact of guidance strength ($\omega$). (b) unconditional dropout probability ($P_{\text{uncond}}$). (c)number of diffusion steps ($T$).}
    \label{fig:parameters}
\end{figure*}

To evaluate the sensitivity of the diffusion-based framework to its key generative hyperparameters, we conducted an ablation study focusing on conditional guidance strength, unconditional dropout probability, and the total number of diffusion steps. The results, presented in Fig. \ref{fig:parameters}(a)–(c), offer valuable insights into the design and implementation of diffusion-based optimization systems for wireless communication tasks.

The guidance strength, denoted as $\omega$, significantly impacts the quality of generation, as illustrated in Fig.~\ref{fig:parameters}(a). Increasing $\omega$ enhances deployment fidelity up to a certain threshold. Beyond this point, however, performance deteriorates due to over-conditioning and noise amplification.

Similarly, Fig.~\ref{fig:parameters}(b) shows that the unconditional dropout probability, $P_{\text{uncond}}$, also affects the model's robustness and adaptability. A low dropout rate (approximately 0.1) improves generalization by encouraging the model to rely on contextual information while preventing overfitting to noisy or incomplete inputs.

The number of diffusion steps, $T$, is another crucial factor, as shown in Fig.~\ref{fig:parameters}(c). Increasing $T$ allows the model to refine its generative trajectory over more iterations, which enhances the quality of the final output. However, the marginal gains in performance diminish after $T = 20$, while computational costs continue to rise. Therefore, $T = 20$ is selected as an empirically effective balance between generation quality and inference efficiency.

\subsection{Generalization Validation}
\begin{figure}
    \centering
\includegraphics[width=1.0\columnwidth]{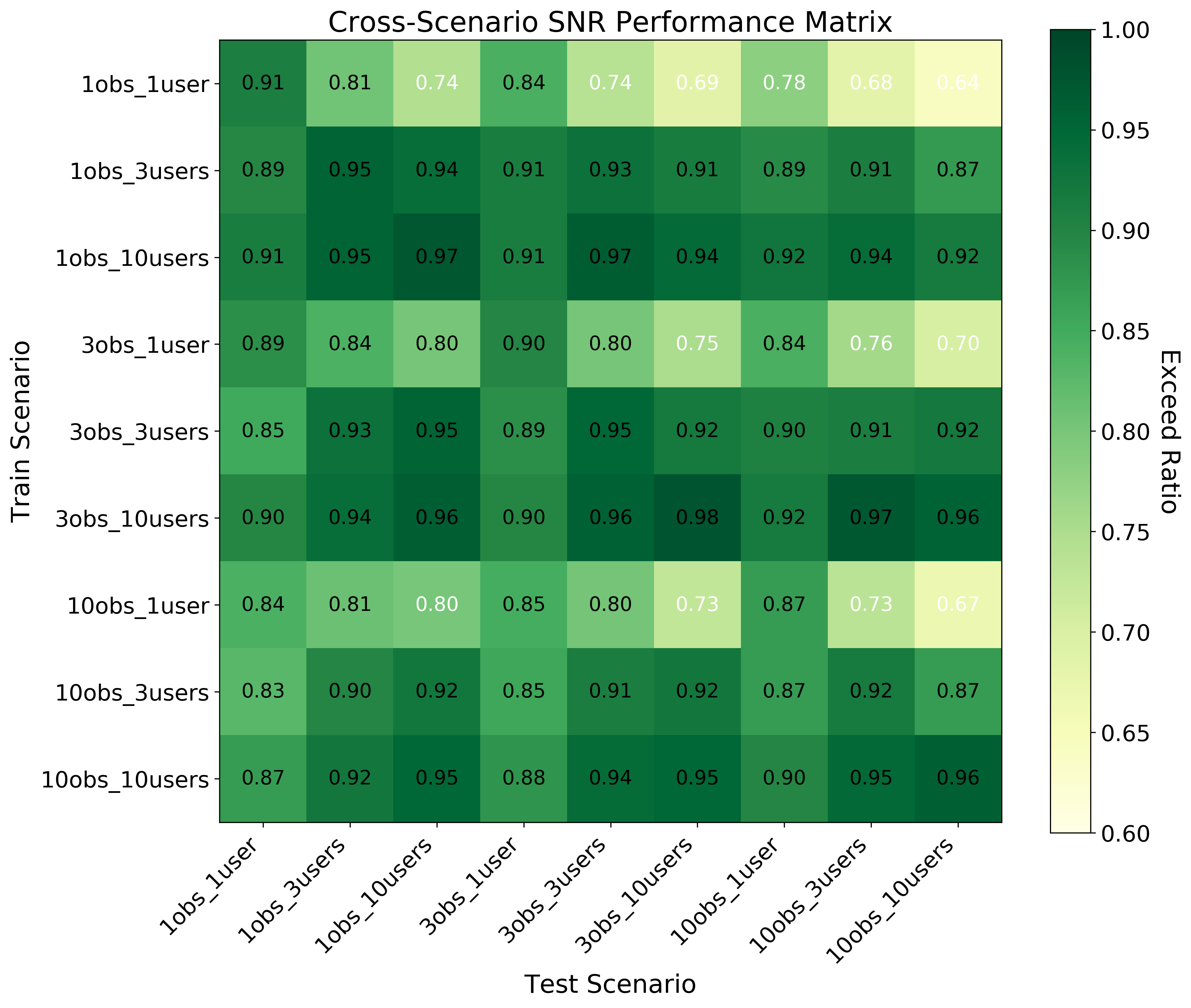}
    \caption{Cross-task generalization results of the proposed diffusion-based RIS deployment method.}
    \label{fig:generalization}
\end{figure}
To further evaluate the generalization ability and robustness of the proposed framework, we conduct experiments across different environments. In each experiment, the diffusion model is trained on a specific dataset (e.g., 3obs$\_$3users) and evaluated on multiple unseen scenarios (e.g., 1obs$\_$3users and 3obs$\_$1user), thereby forming a cross-domain performance matrix. As shown in Fig.~\ref{fig:generalization}, the diffusion model trained on intermediate-complexity scenarios (e.g., 3obs$\_$3users or 3obs$\_$10users) generalizes robustly to both simpler and more complex testing environments. The small differences between diagonal and off-diagonal entries confirm the model's generalization capability. 


\subsection{Performance Comparison}
\begin{figure}
    \centering
    \includegraphics[width=1\linewidth]{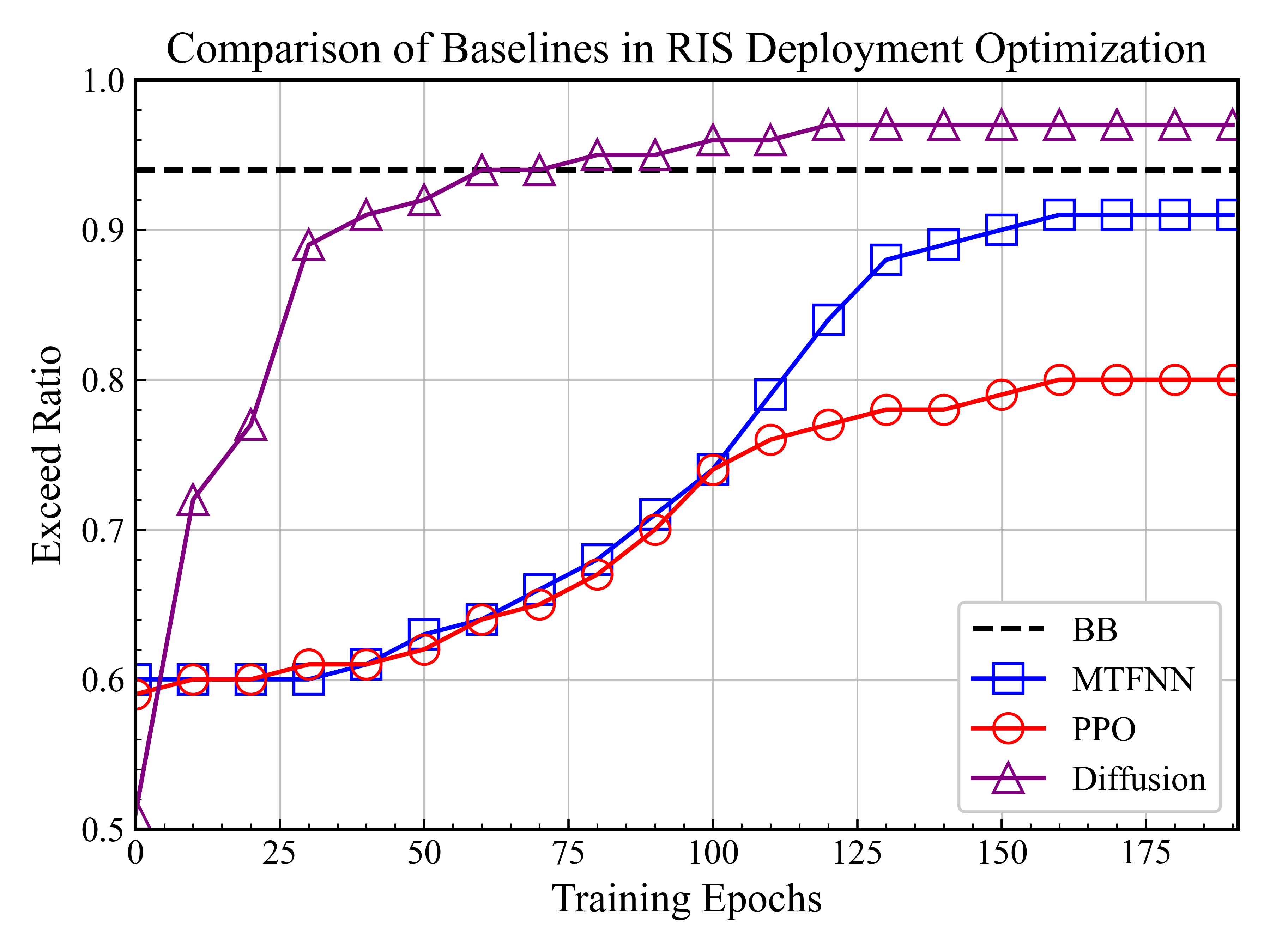}
    \caption{Performance comparison of the proposed diffusion-based RIS deployment method with baselines (BB, PPO, and MTFNN).}
    \label{fig:baselines}
\end{figure}

We compare the proposed diffusion-based framework with three representative baseline methods: (1) branch and bound (BB), a traditional mathematical optimizer; (2) PPO, a reinforcement learning method that is suitable for dynamic tasks but sensitive to reward design; and (3) MTFNN, a supervised predictor that depends on a large amount of labeled data.

All baseline methods are implemented under identical configurations to ensure fairness and consistency. The performance results are summarized in Figure~\ref{fig:baselines}. Among all approaches, the diffusion-based method exhibits the most consistent and superior convergence behavior, achieving an SNR ratio above 0.95 within the first 100 epochs and maintaining near-optimal performance throughout the rest of the training. This finding demonstrates the model’s ability to efficiently approximate the optimal deployment while avoiding local optima and unstable updates. The guided generation mechanism of the diffusion model, along with its probabilistic denoising dynamics, facilitates a smooth trajectory towards optimality, even under complex environmental constraints.



\section{Conclusion}
This paper addresses the critical challenge of optimizing the deployment of RIS in three-dimensional wireless communication environments. By leveraging the generative capabilities of diffusion models, we reframe the deployment task as a conditional generation problem, enabling solutions that are adaptable to dynamic environments. The proposed diffusion-based framework effectively captures the interactions among RIS positions, the locations of the users, BS, and obstacles. Simulation results demonstrate that the proposed framework achieves near-optimal deployment while also improving adaptability in dynamic scenarios. These findings highlight the potential of the generative diffusion model for large-scale and real-time RIS deployment.
\bibliographystyle{IEEEtran}
\bibliography{IEEEabrv,reference}

\end{document}